\documentstyle[multicol,aps,epsf]{revtex}
\begin{document}

\title{
Low-Lying Excitations from the Yrast Line of Weakly Interacting Trapped 
Bosons
}\par

\author{Tatsuya Nakajima$^{1, \ast}$ and Masahito Ueda$^{2}$}\par

\address{$^1$ Department of Physics, Indiana University,
Bloomington, Indiana 47405, USA
}\par

\address{$^2$ Department of Physics, Tokyo Institute of Technology,
Meguro-ku, Tokyo 152-8551, Japan, and \\
CREST, Japan Science and Technology Corporation (JST), Saitama 332-0012, 
Japan
}\par

\date{November 2, 2000, submitted to PRA}

\maketitle

\begin{abstract}
Through an extensive numerical study, we find that the low-lying,
quasi-degenerate eigenenergies of weakly-interacting trapped $N$
bosons with total angular momentum $L$ are given in case of 
small $L/N$ and sufficiently small $L$ by
$E = L\hbar\omega+g[N(N-L/2-1)+1.59\,n(n-1)/2]$,
where $\omega$ is the frequency of the trapping potential and $g$
is the strength of the repulsive contact interaction. The last term
arises from the pairwise repulsive interaction among $n$ octupole
excitations and describes the lowest-lying excitation spectra from
the Yrast line.
In this case, the quadrupole modes do not interact with themselves
and, together with the octupole modes,
exhaust the low-lying spectra that are separated
from others by $N$-linear energy gaps.
\end{abstract}

\pacs{PACS numbers: 03.75.Fi, 05.30.Jp, 67.40.Db}

\begin{multicols}{2}
\narrowtext

The Yrast state, which is a subject of active research in nuclear 
physics~\cite{Hamamoto}, is the lowest-energy state of a system of particles
for a given total angular momentum (AM).
This state is important in that the whole of the excitation energy
is used up for the rotation of the system and hence the system
is at zero temperature, opening up the possibility of performing
precise spectroscopic measurements of energy levels close to the
Yrast state.
Recently, Mottelson~\cite{mot} has pointed out that similar problems
arise in rotating Bose-Einstein condensates (BECs) of trapped atomic
vapor~\cite{review}. Two recent observations of vortices in trapped
BECs~\cite{experi} have lent a great impetus to theoretical studies
of this subject~\cite{mot,thomas,wilk,kavou,jackson,bert,jk,sw,pb}.

The Yrast state is obtained by distributing a given AM $L$ over $N$
bosons so as to minimize the total energy. When the trapping potential
is harmonic, there would be a huge number of such partitions for
$L\gg1$ and hence a huge degeneracy, were it not for the
interactions~\cite{mot}.
The problem thus reduces to finding how a weak interaction lifts the
degeneracy, selects the Yrast state, and determines the low-lying
excitations from this state.
Wilkin {\it et al.}~\cite{wilk} have found that when the interaction is
attractive, all the AM in the Yrast state resides in the
center-of-mass motion of the system.
Mottelson~\cite{mot} discussed that low-lying excitations are well
described by collective modes whose excitation operators are given by
\begin{eqnarray}
\hat{Q}_\lambda=\frac{1}{\sqrt{N \lambda !}}\sum_{p=1}^N
(x_p+iy_p)^\lambda,
\label{col}
\end{eqnarray}
where $\lambda$ is an integer, and $x_p$ and $y_p$ are
the position operators of the
$p$-th particle. The corresponding excitation energy is given by
~\cite{mot}
\begin{equation}
\epsilon _{\lambda} = -2 g N
\bigg (1- \frac{1}{2^{\lambda -1}} \bigg ),
\label {eq: cee}
\end{equation}
where $g$ denotes the strength of the contact interaction.            
Equation~(\ref {eq: cee}) suggests that for the case of attractive    
interactions ({\it i.e.}, $g<0$), the AM of the Yrast state is carried
by the dipolar ($\lambda=1$) mode, in agreement with Ref.~\cite{wilk},
while for the case of repulsive interactions it is carried by the     
quadrupole ($\lambda=2$) or octupole ($\lambda=3$) modes \cite{mot},  
as they have the greatest energy gain per unit of AM. A more          
elaborate study taking the mode-mode interaction into account has     
shown that the quadrupole modes have slightly larger energy gains than
the octupole modes~\cite{kavou}.

Bertsch and Papenbrock~\cite{bert} have performed numerical diagonalization
for small systems ($N=25$ and $50$), finding that the energy of the
Yrast line for $L\leq N$ can be given by $E=L\hbar\omega+gN(N-L/2-1)$,  
where $\omega$ is the frequency of the trapping potential.  The     
corresponding eigenstate was very recently shown to exist~\cite{jk} 
(see also Refs.~\cite{sw,pb}). This state partitions the AM equally 
among the particles~\cite{bert,sw,pb} and smoothly crosses over to  
the many-body single vortex state~\cite{wilk} as $L$ approaches $N$.

In this paper, we report on the results of our extensive numerical  
study of this system with up to 640000 particles and for a total AM 
up to 30.
Our primary finding is that of the {\it lowest-lying} excitation
spectra from the Yrast line that arise from the pairwise repulsive
interaction between octupole modes; the corresponding interaction
energy is given by $1.59\,gn(n-1)/2$, 
where $n$ is the number of
excited octupole modes~\cite{un1}. 
These energy levels are separated from the other excitations, including
the one discussed in Ref.~\cite{kavou}, by $N$-linear energy gaps.
We have also found that 
for small $L/N$ and sufficiently small $L$ 
all the many-body eigenstates have
surprisingly large overlap ($\sim 0.99$) with trial wavefunctions
defined by
\begin{eqnarray}
\prod_{\lambda=1}^L(\hat{Q}_{\lambda})^{n_{\lambda}}|0 \rangle
\ \ {\rm with} \ \sum_{\lambda = 1}^L \lambda n_{\lambda} = L,
\label{variational}
\end{eqnarray}
where $n_{\lambda}$ is a non-negative integer and
$|0\rangle$ describes the exact many-body ground state of the
system with $L=0$.

The model we study is the same as that of 
Refs.~\cite{wilk,kavou,jackson,bert}.
We consider a system of weakly-interacting bosons subject to a given
AM and trapped in a parabolic confining potential. The system is
isotropic in the $x-y$ plane and strongly confined in the $z$
direction so that all the particles occupy the lowest-energy state
in this direction.
The problem is thus essentially two-dimensional.
The many-body Hamiltonian for this system can be written as
$\hat{H} = \sum_i \hat{h}_i + \hat{V}$. Here $\hat{h}_i$ denotes the
single-particle Hamiltonian for the $i$-th particle
\begin{equation}
\hat {h}_i = - \frac {\hbar ^2}{2M}
\left (\frac {\partial ^2 \ }{\partial x_i^2}
+\frac {\partial ^2 \ }{\partial y_i^2} \right )
+ \frac {M \omega ^2}{2} (x_i^2+y_i^2),
\label{single}
\end{equation}
where $M$ is the mass of the boson,
$\omega$ is the frequency of the trap, and
$\hat{V} = (4 \pi \hbar ^2 \tilde {g}/M) \,
\sum _{i<j}
\delta ^{(2)} ({\bf r}_i - {\bf r}_j) $
describes the interaction between particles ($\tilde {g}$ is the
dimensionless `scattering length').
In this paper we consider the
case of $\tilde {g} > 0$, {\it i.e.}, repulsive interactions.

The single-particle spectrum is given by
$E_{n,m} = \hbar \omega \,(2n+|m|+1)$, where $n$ and $m$ denote the
radial and AM quantum numbers, respectively. When only low-lying
states of the many-boson system are concerned, we may assume that all
the particles are in states with $n = 0$. The single-particle states
are then labeled only by $m$ and described by
$\phi _m (z) = (z^m/\sqrt{\pi m!}) \exp (- |z|^2/ 2)$,
where $z \equiv x + i y$, and the lengths are measured in units of
$(\hbar/M \omega)^{1/2}$. We use these single-particle states as a
basis set to rewrite the many-body Hamiltonian as
\begin{eqnarray}
\hat {H} &=& \hbar \omega \, \sum _{m \geq 0}
m \,b^{\dagger}_m b_m \nonumber\\
& &+ g \sum _{m_1 \sim m_4}
V_{m_1 m_2 m_3 m_4} \,
b^{\dagger}_{m_1} b^{\dagger}_{m_2} b_{m_3} b_{m_4},
\label {eq: hammd}
\end{eqnarray}
where $g \equiv \tilde {g} \hbar \omega$, the operators $b_m$ and
$b^{\dagger}_m$ annihilate and create one boson in the single-particle
state $\phi _m$, respectively, and $V_{m_1 m_2 m_3 m_4}$ is the
two-body matrix element given as~\cite{bert}
\begin{equation}
V_{m_1 m_2 m_3 m_4} =
\frac {\delta _{m_1+m_2, m_3+m_4} \,(m_1+m_2)!}
{2^{m_1+m_2}\,\sqrt{m_1 !\,m_2 !\,m_3 !\,m_4 !}} .
\label {eq: hammd1}
\end{equation}
Given a total AM $L$, the state may be spanned by the Fock states
$|n_0, n_1, \ldots, n_L \rangle$ with $\sum _{i=0,L} n_i = N$ and
$\sum_{j=0,L}j n_j = L$, where $n_i$ denotes the occupation number of
the $i$-th single particle state $\phi _i$, {\it i.e.},
$b_i^\dagger b_i |
n_0, n_1, \ldots, n_L \rangle=n_i |n_0, n_1,\ldots, n_L \rangle$.
Numerical diagonalization of the two-body interaction $\hat{V}$ is
performed within the Hilbert subspace subject to these constraints.

The energy eigenvalue corresponding to the $i$-th many-body eigenstate
with total AM $L$ may be written as $L \hbar\omega + \epsilon_{L, i}$,
where $\epsilon _{L, i}$ denotes the interaction energy. Because the
lowest-energy eigenvalue with total AM  $L$ ($2 \le L \le N$) is
given by $\epsilon_{L} = g N (N-L/2-1)$, the unique state with $L=0$,
which is composed of $N$ bosons in a single-particle state $\phi _0$,
is the ground state for $g N \ll \hbar \omega$, and it will therefore
be denoted as $| 0 \rangle$. Our trial wavefunctions are constructed
by acting on the ground state $| 0 \rangle$ with the collective
operators, $\{ \hat {Q}_{\lambda} \}$, whose second quantized forms
are given by $\hat {Q}_{\lambda} = (1/\sqrt{N}) \sum _{m}
\sqrt {(m + \lambda) !/m ! \,\lambda ! } \,
b^{\dagger}_{m + \lambda} b_{m}$.


\begin{figure}[t]
\begin{center}
\leavevmode\epsfxsize=68mm \epsfbox{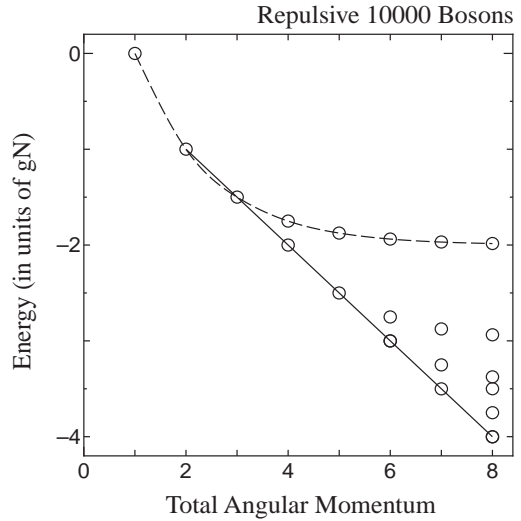}
\end{center}
\vspace*{-1.0cm}
\caption{The excitation spectra for $N=10000$, $L \leq 8$.
The energy is measured in units of $gN$, and the Yrast line
is shown as a solid line.
The collective excitations,
$\hat {Q}_{\lambda} | 0 \rangle$ ($\lambda \leq 8$),
are linked by a dashed curve,
and their energy values are given in Table I.
Except for the state $\hat {Q}_1 | 0 \rangle$,
the states that include excitations of 
the center-of-mass motion are omitted.
Each state has a very large overlap ($\geq 0.994$)
with the corresponding trial wavefunction.
These states are given in the ascending order of energy as
$\hat {Q}_1 |0 \rangle$ for $L=1$;
$\hat {Q}_2 |0 \rangle$ for $L=2$;
$\hat {Q}_3 |0 \rangle$ for $L=3$;
$(\hat {Q}_2)^2 |0 \rangle$, $\hat {Q}_4 |0 \rangle$ for $L=4$;
$\hat {Q}_2 \hat {Q}_3 |0 \rangle$,
$\hat {Q}_5 |0 \rangle$ for $L=5$;
$(\hat {Q}_2)^3 |0 \rangle$, $(\hat {Q}_3)^2 |0 \rangle$,
$\hat {Q}_2 \hat {Q}_4 |0 \rangle$,
$\hat {Q}_6 |0 \rangle$ for $L=6$;
$(\hat {Q}_2)^2 \hat {Q}_3 |0 \rangle$,
$\hat {Q}_3 \hat {Q}_4 |0 \rangle$,
$\hat {Q}_2 \hat {Q}_5 |0 \rangle$,
$\hat {Q}_7 |0 \rangle$ for $L=7$;
$(\hat {Q}_2)^4 |0 \rangle$, $\hat {Q}_2 (\hat {Q}_3)^2 |0 \rangle$,
$(\hat {Q}_2)^2 \hat {Q}_4 |0 \rangle$,
$(\hat {Q}_4)^2 |0 \rangle$,
$\hat {Q}_3 \hat {Q}_5 |0 \rangle$,
$\hat {Q}_2 \hat {Q}_6 |0 \rangle$,
$\hat {Q}_8 |0 \rangle$ for $L=8$.
The sets, [$(\hat {Q}_2)^3 |0 \rangle$, $(\hat {Q}_3)^2 |0 \rangle$]
for $L=6$
and [$(\hat {Q}_2)^4 |0 \rangle$,
$\hat {Q}_2 (\hat {Q}_3)^2 |0 \rangle$] for $L=8$,
are quasi-degenerate,
and these states are shown on
or extremely near the Yrast line.
}
\label{fig1}
\end{figure}

Figure~\ref{fig1} shows the energy spectra for $N=10000$ and $L\leq8$,
where the energy is measured in units of $gN$. Our numerical results
show that all the energy eigenstates for $L \leq 8$ have remarkably
large overlaps ($\geq 0.994$) with the corresponding trial
wavefunctions given in Eq.~(\ref{variational}). Among the eigenstates,
only ``intrinsic" ones, {\it i.e.},
those that do not include dipole
excitations, are shown in Fig.~\ref{fig1} (the only exception being
the state $\hat {Q}_1 | 0 \rangle$ for $L=1$),
as the dipole mode corresponds to the center-of-mass motion that does 
not affect the interaction in a harmonic trap.
For example, for $L=6$ only the ``intrinsic" eigenstates,
$(\hat {Q}_2)^3 |0 \rangle$, $(\hat {Q}_3)^2 |0 \rangle$,
$\hat {Q}_2 \hat {Q}_4 |0 \rangle$, and
$\hat {Q}_6 |0 \rangle$, are shown
among all the eigenstates for $L=6$, and they 
are identified in the ascending order of energy as
$(\hat {Q}_2)^3 |0 \rangle$,
$(\hat {Q}_3)^2 |0 \rangle$,
$\hat {Q}_2 \hat {Q}_4 |0 \rangle$,
$\hat {Q}_1 \hat {Q}_2 \hat {Q}_3 |0 \rangle$,
$(\hat {Q}_1)^2 (\hat {Q}_2)^2 |0 \rangle$,
$\hat {Q}_6 |0 \rangle$, $\hat {Q}_1 \hat {Q}_5 |0 \rangle$,
$(\hat {Q}_1)^2 \hat {Q}_4 |0 \rangle$,
$(\hat {Q}_1)^3 \hat {Q}_3 |0 \rangle$,
$(\hat {Q}_1)^4 \hat {Q}_2 |0 \rangle$,
$(\hat {Q}_1)^6 |0 \rangle$.
We should note here that two states, 
$(\hat{Q}_2)^3|0\rangle$ and $(\hat{Q}_3)^2|0 \rangle$, are seen 
as `a single circle' in Fig.~\ref{fig1} due to
their {\it quasi-degeneracy}. The energy difference between them is
caused by mode-mode interactions 
and is on the order of $g$~\cite{mot,kavou,pb}. Two states
$(\hat {Q}_2)^4 |0 \rangle$ and $\hat {Q}_2 (\hat {Q}_3)^2 |0\rangle$
for $L=8$ are also quasi-degenerate. The Yrast line is shown by a
solid line in Fig.~\ref{fig1}, and the other states are separated
from this line by energy gaps roughly equal to or greater than $gN/4$. The
energy difference $gN/4$ corresponds to the replacement of the factor
$(\hat {Q}_2)^2$ included in the Yrast states with $\hat {Q}_4$.

In Fig.\ref{fig1}, the collective excitations discussed by Mottelson,
$\hat {Q}_{\lambda} | 0 \rangle$ ($\lambda = 1,2,\ldots,8$), are
linked by a dashed curve. The energies of these states are given in
Table \ref{tab1}. These values are very well described by
Eq.~(\ref {eq: cee}); in particular, for $\lambda \leq 3$ the
agreement is perfect within the machine accuracy~\cite{bert}.

\begin{figure}[t]
\vspace*{-1.0cm}
\begin{center}
\leavevmode\epsfxsize=68mm \epsfbox{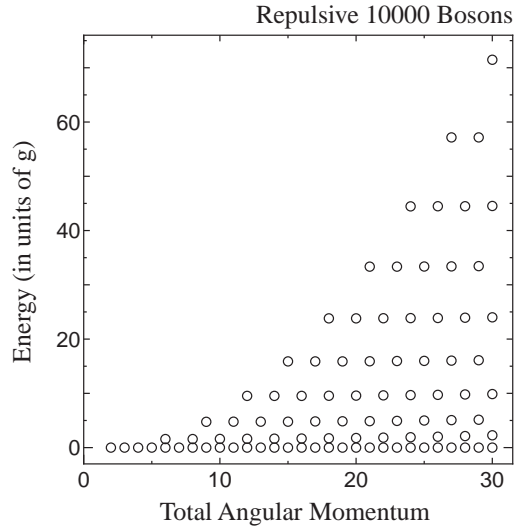}
\end{center}
\caption{The low-lying, quasi-degenerate eigenstates
for $N=10000$, $2 \leq L \leq 30$ are shown by open circles.
These states are well described by the
quadrupole and octupole excitation modes, as seen in Table II.
These energy values, which are measured from the Yrast line,
are well described by $1.59 \times n(n-1)/2$ in units of $g$.
Here, $n$ is the number of excited octupole modes.
As the total AM increases,
the interconversion between these two modes
becomes apparent (see Table V),
which is also suggested by the level repulsion between
the lowest and first-excited states for larger $L$.
}
\label{fig2}
\end{figure}

For higher (but not very large) AM, 
the low-lying states are again well described by the
collective excitations. 
In particular, the quadrupole ($\lambda = 2$)
and octupole ($\lambda = 3$) modes well describe the low-lying,
quasi-degenerate states as shown in Table \ref{tab2}. These
quasi-degenerate states for $2 \leq L \leq 30$ are shown in
Fig.~\ref{fig2}, where the energies are measured from the Yrast line
and are on the order of $g$. We note that the other excitations are
separated from these in Fig.~2 by $N$-linear energy gaps ($\geq gN/4$).

However,
the overlap integrals between the collective trial wavefunctions 
and the numerically-obtained low-lying eigenstates become small 
even for small $L/N$ as $L$ increases.
In fact, in Table \ref{tab3new}, 
the deviations of these overlaps from unity
seem to include not only a factor of $L/N$ but also another
factor like $L ^s$ ($s \sim 2$) for small $L$.
Thus these trial wavefunctions describe the low-lying states well
for small $L/N$ and sufficiently small $L$.

Moreover, for small $L/N$ and sufficiently small $L$, 
the energies of these quasi-degenerate 
states are found to be described by $1.59\,g \times n(n-1)/2$, where
$n$ is the number of excited octupole modes in each state \cite{un1,un2}.
This energy `quantization' can also be seen in the data for
$N=640000$, $L=18$ in Table \ref{tab4} as $1.590 \simeq 1.59 \times
2(2-1)/2$, $9.530 \simeq 1.59 \times 4(4-1)/2$ and $23.82 \simeq 1.59
\times 6(6-1)/2$. Here, $2$, $4$ and $6$ are the number of excited
octupole modes in the first, second, and third excited states for
$L=18$, as seen in Table \ref{tab2}.

These results show that the quadrupole modes do not interact with
themselves or with the octupole mode in the limit of
$L/N \rightarrow 0$, while the octupole modes undergo a pairwise
repulsive interaction. 
Although the mode-mode interactions were discussed in 
Refs.~\cite{mot,kavou,pb}, such concrete quasiparticle features of 
these modes were not pointed out previously.
Because of such quasiparticle features, 
the Yrast states are well described by $(\hat {Q}_2)^n | 0
\rangle$ for even $L$ and by $(\hat {Q}_2)^{n^{\prime}} \hat {Q}_3 |0
\rangle$  for odd $L$ in that limit ($n$, $n^{\prime}$: integers).
The lowest energy for sufficiently small AM $L$
is therefore simply given by the sum of the excitation energies of 
the quadrupole modes (and the energy of an octupole mode for odd $L$).

As the total AM increases, however, the interconversion between these
two modes becomes significant. This is seen in Table \ref{tab5} and
is also suggested by the level repulsion between the lowest and
first-excited states for larger $L$ in Fig.~\ref{fig2}. Although 
contributions by the modes, $\hat {Q}_4$, $\hat {Q}_5$ and $\hat {Q}_6$
to the Yrast states have been claimed~\cite{kavou},
the interconversion between 
modes $\hat{Q}_2$ and $\hat{Q}_3$ makes a greater contribution than
these three modes, as seen in Table~\ref{tab5}.

Whereas the Yrast state is analytically given for $L \leq N$
\cite{sw,pb}, its
relations to collective excitations 
for small $L/N$ and sufficiently small $L$
(especially the
interconversion effects between the quadrupole and octupole modes)
remain unclear. An understanding of these relations may help us 
clarify the integrability in a two-dimensional bose system with a 
 harmonic potential that includes weak repulsive delta-function   
interactions or a hidden symmetry of the Hamiltonian~\cite{pita}.


T.N. was supported for the research in the USA by 
Japan Society for the Promotion of Science.
M.U. acknowledges support by a Grant-in-Aid for Scientific Research
(Grant No. 11216204) by the Ministry of Education, Science, Sports,
and Culture of Japan, and by the Toray Science Foundation.



\end{multicols}
\widetext

\vspace*{1.0cm}

\begin{table}
\begin{tabular}[t]{||c|c|c|c|c|c|c|c|c||}
$\lambda$ & 1 & 2 & 3 & 4 & 5 & 6 & 7 & 8 \\
\hline
$\epsilon_{\lambda}/gN$ & 0 & -1 & -3/2 & -1.7499 & -1.8749
& -1.9373 & -1.9686 & -1.9842 
\end{tabular}
\caption{
Energies of the collective modes that are excited by acting
on the ground state of $N=10000$ bosons with
the excitation operator $\hat{Q}_{\lambda}$.
The agreement with eq.~(2) is perfect (within
the machine accuracy) for $\lambda \leq 3$ and excellent for
$\lambda \geq 4$.
}
\label{tab1}
\end{table}

\begin{table}
\begin{tabular}{||c|c|c|c|c|c|c||}
 & {\rm lowest} & 1{\rm st-excited} & 2{\rm nd-excited}
& 3{\rm rd-excited} & 4{\rm th-excited} & 5{\rm th-excited} \\
\hline
$L=9$ & 0.9951 & 0.9977 & & & & \\
 & $(\hat {Q}_2)^{3} \hat {Q}_3 |0 \rangle$
& $(\hat {Q}_3)^{3} |0 \rangle$ & & & & \\
\hline
12 & 0.9804 & 0.9815 & 0.9959 & & & \\
  & $(\hat {Q}_2)^{6} |0 \rangle$
& $(\hat {Q}_2)^{3} (\hat {Q}_3)^{2} |0 \rangle$
& $(\hat {Q}_3)^{4} |0 \rangle$ & & & \\
\hline
15 & 0.9820 & 0.9846 & 0.9936 & & & \\
 & $(\hat {Q}_2)^{6} \hat {Q}_3 |0 \rangle$
& $(\hat {Q}_2)^{3} (\hat {Q}_3)^{3} |0 \rangle$
& $(\hat {Q}_3)^{5} |0 \rangle$ & & & \\
\hline
18 & 0.9374 & 0.9363 & 0.9812 & 0.9906 & & \\
  & $(\hat {Q}_2)^{9} |0 \rangle$
& $(\hat {Q}_2)^{6} (\hat {Q}_3)^{2} |0 \rangle$
& $(\hat {Q}_2)^{3} (\hat {Q}_3)^{4} |0 \rangle$
& $(\hat {Q}_3)^{6} |0 \rangle$ & & \\
\hline
30 & 0.7674 & 0.7447 & 0.9105 & 0.9418 & 0.9593 & 0.9735 \\
  & $(\hat {Q}_2)^{15} |0 \rangle$
& $(\hat {Q}_2)^{12} (\hat {Q}_3)^{2} |0 \rangle$
& $(\hat {Q}_2)^{9} (\hat {Q}_3)^{4} |0 \rangle$
& $(\hat {Q}_2)^{6} (\hat {Q}_3)^{6} |0 \rangle$
& $(\hat {Q}_2)^{3} (\hat {Q}_3)^{8} |0 \rangle$
& $(\hat {Q}_3)^{10} |0 \rangle$ 
\end{tabular}
\caption{
Values of the overlap integrals between the numerically found
low-lying quasi-degenerate states with $N=10000$ for various
values of $L$ and the corresponding normalized trial
wavefunctions (shown on the second line in each cell).
These values are close to unity for small $L/N$ and 
sufficiently small $L$.
}
\label{tab2}
\end{table}


\begin{table}
\begin{tabular}{||c|c|c|c|c|c|c|c||} 
          & $L=6$ & 8 & 10 & 12 & 14 & 16 & 18 \\
\hline
 $N=5000$ & 0.0049 & 0.0117 & 0.0227 & 0.0385 
 & 0.0596 & 0.0861 & 0.1179 \\
\hline
 $10000$ & 0.0025 & 0.0059 & 0.0115 & 0.0196 
 & 0.0307 & 0.0450 & 0.0626 \\
\hline
 $20000$ & 0.0012 & 0.0030 & 0.0058 & 0.0099 
 & 0.0156 & 0.0230 & 0.0323 
\end{tabular}
\caption{The deviations of overlap integrals from unity 
are shown for $N=5000,10000,20000$ and $L=6,8,10,12,14,16,18$,
where each overlap integral is the one between a normalized trial 
wavefunction with $n$ quadrupole excitations 
({\it i.e.}, $\propto (\hat {Q}_2)^{n} |0 \rangle$\,) and 
a numerically-obtained lowest-energy state for $L=2n$ ($n$: integer).
These values seem to include not only a factor of $L/N$ but 
also another factor like $L ^s$ ($s \sim 2$) for small $L$.}
\label{tab3new}
\end{table}

\begin{table}
\begin{tabular}{||c|c|c|c|c||}
& $N=10000$ & 40000 & 160000 & 640000 \\
\hline
1{\rm st-excited} &  1.721    &  1.622    &  1.597    &  1.590  \\
2{\rm nd-excited} &  9.564    &  9.538    &  9.532    &  9.530  \\
3{\rm rd-excited} &  23.82    &  23.82    &  23.82    &  23.82  
\end{tabular}
\caption{
Energies (in units of $g$) of the low-lying quasi-degenerate states
with $L=18$ for various values of $N$. They are measured from the
Yrast line.  As the ratio $L/N$ becomes small, the convergence of
these energies can be seen.
}
\label{tab4}
\end{table}

\begin{table}
\begin{tabular}{||c|c|c|c|c|c||}
&
$(\hat {Q}_2)^{15} |0 \rangle$ &
$(\hat {Q}_2)^{12} (\hat {Q}_3)^{2} |0 \rangle$ &
$(\hat {Q}_2)^{9} (\hat {Q}_3)^{4} |0 \rangle$ &
$(\hat {Q}_2)^{13} \hat {Q}_4 |0 \rangle$ &
$(\hat {Q}_2)^{10} (\hat {Q}_3)^{2} \hat {Q}_4 |0 \rangle$ \\
\hline
{\rm lowest}      & 0.7674 & 0.5307 & 0.1045
& 0.1338 & 0.0737  \\
1{\rm st-excited} & 0.5400 & 0.7447 & 0.1928
& 0.0937 & 0.1035 \\
\end{tabular}
\caption{
Values of the overlap integrals between the lowest/first-excited
states and some trial wavefunctions for $N=10000$ and $L=30$.
The interconversion between the quadrupole and octupole modes
makes a much larger contribution to the Yrast state than other
modes such as $\hat {Q}_4$.
}
\label{tab5}
\end{table}


\begin{thebibliography}{99}


\bibitem[\ast]{a} Present address: Department of Physics, 
University of Texas, Austin, Texas 78712.
Permanent address:
Physics Department, Graduate School of Science, Tohoku University,
Sendai 980-8578, Japan. 

\bibitem{Hamamoto} I. Hamamoto, in {\it Treatise on Heavy-Ion Science},
edited by D. A. Bromley (Plenum, New York, 1985), vol.3, p.313;
I. Hamamoto and B. Mottelson, Nucl. Phys. {\bf A507}, 65 (1990).

\bibitem{mot} B. Mottelson,
Phys. Rev. Lett. {\bf 83}, 2695 (1999).

\bibitem{review} F. Dalfovo, S. Giorgini, L.P. Pitaevskii
and S. Stringari, Rev. Mod. Phys. {\bf 71}, 463 (1999).

\bibitem{experi} M.R. Matthews, B.P. Anderson, P.C. Haljan, D.S. Hall,
C.E. Wieman and E.A. Cornell, Phys. Rev. Lett. {\bf 83}, 2498 (1999);
K.W. Madison, F. Chevy, W. Wohlleben and J. Dalibard,
{\it ibid.} \,{\bf 84}, 806 (2000); preprint (cond-mat/0004037).

\bibitem{thomas} D.A. Butts and D.S. Rokhsar,
Nature {\bf 397}, 327 (1999);
M. Linn and A.L. Fetter, preprint (cond-mat/9906139);
D.L. Feder, C.W. Clark and B.I. Schneider,
preprint (cond-mat/9910288); preprint (cond-mat/9904269).

\bibitem{wilk} N.K. Wilkin, J.M.F. Gunn and R.A. Smith,
Phys. Rev. Lett. {\bf 80}, 2265 (1998).

\bibitem{kavou} G.M. Kavoulakis, B. Mottelson
and C.J. Pethick, preprint (cond-mat/0004307).

\bibitem{jackson} A.D. Jackson, G.M. Kavoulakis, B. Mottelson
and S.M. Reimann, preprint (cond-mat/0004309).

\bibitem{bert} G.F. Bertsch and T. Papenbrock,
Phys. Rev. Lett. {\bf 83}, 5412 (1999).

\bibitem{jk} A.D. Jackson and G.M. Kavoulakis,
preprint (cond-mat/0005159).

\bibitem{sw} R.A. Smith and N.K. Wilkin,
preprint (cond-mat/0005230).

\bibitem{pb} T. Papenbrock and G.F. Bertsch,
preprint (cond-mat/0005480).

\bibitem{un1} The pairwise interaction term between octupole 
modes can also be given analytically as
$(27/17) \times n (n-1)/2$ [M. Ueda and T. Nakajima, unpublished].

\bibitem{un2}
This mode-mode interaction term is obtained by 
considering the energy contribution 
by four diagrams in Fig.~5 in Ref.~\cite{kavou} 
up to the order of $g$.

\bibitem{pita} L.P. Pitaevskii and A. Rosch,
Phys. Rev. A {\bf 55}, R853 (1997).


\end{thebibliography}
\end{document}